# Approximating (Unweighted) Tree Augmentation via Lift-and-Project Part 0: $(1.8 + \epsilon)$ approximation for (Unweighted) TAP

Joseph Cheriyan    Zhihan Gao

7 July 2014

**Abstract**

We study the unweighted *Tree Augmentation Problem* (TAP) via the Lasserre (Sum of Squares) system. We prove an approximation guarantee of $(1.8 + \epsilon)$ relative to an SDP relaxation, which matches the combinatorial approximation guarantee of Even, Feldman, Kortsarz and Nutov in ACM TALG (2009), where $\epsilon > 0$ is a constant. We generalize the combinatorial analysis of integral solutions of Even, et al., to fractional solutions by identifying some properties of fractional solutions of the Lasserre system via the decomposition result of Rothvoß (arXiv:1111.5473, 2011) and Karlin, Mathieu and Nguyen (IPCO 2011).

This is a manuscript from July 2014 that was widely circulated but was not posted on the web. It is being posted for the sake of archiving/referencing. The results have been subsumed by arXiv:1507.01309, our manuscript from July 2015, that proves an approximation guarantee of $(1.5 + \epsilon)$ for (unweighted) TAP via a relaxation.

The presentation of the weaker result here may be of interest because it is significantly shorter and simpler than arXiv:1507.01309.

To the best of our knowledge, this was the earliest manuscript to prove an upper-bound less than 2 for the integrality ratio for a relaxation of TAP. (Although approximation guarantees less than 2 for (unweighted) TAP had been published earlier, those guarantees were proved w.r.t. an integral optimal solution and not w.r.t. a relaxation.)

The original manuscript also had a proof of a $(1.75 + \epsilon)$ approximation guarantee but that part has been omitted because the details are difficult; that result has been subsumed by arXiv:1507.01309.

All our results use the same relaxation, namely, a Lasserre tightening of a simple LP relaxation.

# APPROXIMATING (UNWEIGHTED) TREE AUGMENTATION VIA LIFT-AND-PROJECT

JOSEPH CHERIYAN AND ZHIHAN GAO

ABSTRACT. We study the unweighted *Tree Augmentation Problem* (TAP) via the Lasserre (Sum of Squares) system. We prove an approximation guarantee of $(1.8 + \epsilon)$ relative to an SDP relaxation, which matches the combinatorial approximation guarantee of Even, Feldman, Kortsarz and Nutov in ACM TALG (2009), where $\epsilon > 0$ is a constant. We generalize the combinatorial analysis of integral solutions of Even, et al., to fractional solutions by identifying some properties of fractional solutions of the Lasserre system via the decomposition result of Rothvoß (arXiv:1111.5473, 2011) and Karlin, Mathieu and Nguyen (IPCO 2011).

1. INTRODUCTION

2. PRELIMINARIES AND NOTATION

We adopt the notation and terms of Even, et al., [6], where possible; this will aid readers familiar with that paper.

Let $G = (V, E(G))$ be a connected, undirected graph, and let $T = (V, \widehat{E}_T)$ be a spanning tree of $G$. By a *tree-edge* we mean an edge of $T$. Let $E$ denote the edge-set $E(G) - \widehat{E}_T$; we call $E$ the *link set* and we call an element $e \in E$ a *link*; thus, a link is an edge of $G$ that can be used to augment $T$. An instance of TAP consists of $G$ and $T$. We assume that all instances of interest have feasible solutions, that is, we assume that $(V, \widehat{E}_T \cup E)$ is 2-edge connected. The goal is to find a minimum-size subset $F$ of $E$ such that augmenting $T$ by $F$ results in a 2-edge connected graph, i.e., the graph $(V, \widehat{E}_T \cup F)$ is 2-edge connected. (Remark: usually, we use $J$ to denote an arbitrary subset of $E$.)

One of the nodes $r$ of $T$ is designated as the root; thus, we have a rooted tree $T, r$.

For two nodes $i, j \in V$, we use $P_{i,j} = P_{j,i}$ to denote the unique path of the tree $T$ between $i$ and $j$.

The *height* $h(i)$ of a node $i$ in $T$ is defined to be the length of the path $P_{i,r}$.

Let $i$ be a node of $T$. If another node $j$ belongs to the path $P_{i,r}$, then $j$ is called an *ancestor* of $i$, and $i$ is called a *descendant* of $j$. If a descendant $j$ of $i$ is adjacent to $i$ (thus, $j \neq i$), then $j$ is called a *child* of $i$, and $i$ is called a *parent* of $j$. Clearly, every node (except $r$) has a unique parent. If node $i$ has no child, then we call $i$ a *leaf*. Note that $r$ is not a leaf. Throughout, we use $L$ to denote the set of leaves; thus, $L = \{v \in V : v \text{ is a leaf of } T\}$.

For any node $i$, we use $T_i$ to denote the rooted subtree of $T, r$ induced by $i$ and its descendants.

For any $U \subseteq V$, we denote the set of links with both end-nodes in $U$ by $E(U)$, and for any two subsets $U, W$ of $V$, we denote the set of links with one end-node in $U$ and the other end-node in $W$ by $E(U, W)$; thus, $E(U, W) \equiv \{uw \in E : u \in U, w \in W\}$.

For any node $v \in V$, we use $\delta_E(v)$ to denote the set of links incident to $v$, thus, $\delta_E(v) = \{iv : iv \in E\}$.

We say that a link $uw$ *covers* a tree-edge $\hat{e}$ if $P_{u,w} \ni \hat{e}$. Similarly, we say that a rooted subtree $T_i$ is *covered* by a set of links $J \subseteq E$ if each tree-edge of $T_i$ is covered by some link of $J$.

For any tree-edge $\hat{e} \in \widehat{E}_T$, we use $\hat{\delta}_E(\hat{e})$ to denote the set of links that cover $\hat{e}$, thus, $\hat{\delta}_E(\hat{e}) = \{ij \in E : \hat{e} \in P_{i,j}\}$.

For any leaf $v$, we use $\tilde{P}_v$ to denote the path in $T$ between $v$ and either its closest ancestor with more than one child or $r$; thus, $\tilde{P}_v$ is a maximal path whose internal nodes are ancestors of $v$ with exactly one child.

We call a node $s$ of $T$ a *stem* if $s$ has exactly two children, $s$ has exactly two descendants that are leaves, and there exists a link in $E$ between the two leaves of $T_s$; we call the link between the two leaves of $T_s$ a *twin link*; this differs slightly from [6, Definition 3.1]. Observe that there is a one-to-one corresponding between twin links and stems. We may assume that the root $r$ is not a stem, otherwise, we have a trivial instance of TAP. Throughout, we use $\mathcal{S}$ to denote the set of stems; thus, $\mathcal{S} = \{v \in V : v \text{ is a stem of } T\}$. Moreover, we use $\mathcal{R}$ to denote the set of nodes that are neither stems nor leaves; thus $\mathcal{R} = V - (\mathcal{S} \cup L)$. Consider a stem $s$ and let $i, j$ be the two leaves of $T_s$; observe that $T_s$ is the union of $\tilde{P}_i$ and $\tilde{P}_j$. (The notion of stems and twin links is due to [7].)

Let $E_{twin}$ denote the set of twin links, and let $E_{ntw}$ denote the set of leaf-to-leaf links that are not twin links; thus, $E_{ntw} = E(L) - E_{twin}$.

For two links $i_1 j_1$ and $i_2 j_2$, if $P_{i_1 j_1} \subseteq P_{i_2 j_2}$, then we call $i_1 j_1$ a *shadow* (or, *sublink*) of $i_2 j_2$. In particular, if $P_{i_1 j_1} \subsetneq P_{i_2 j_2}$, then we call $i_1 j_1$ a *proper shadow* (or, *proper sublink*) of $i_2 j_2$.

For any link $ij \in E$, if all sublinks of $ij$ also exist in $E$, then we call $E$ *shadow closed*. Clearly, if $E$ is not shadow closed, then we can make it shadow closed by adding all sublinks of each of the original links. It can be seen that this preserves the optimal value of any instance of TAP. Following Event, at al., [6], we make the next assumption (see Assumption 2.2 of [6]).

**Assumption:** $E$ is shadow-closed.

For any leaf $v$ of $T$, $up(v)$ denotes a node $q$ of minimum height that is adjacent to $v$ via a link, thus, $vq \in E$ and $h(q) = \min\{h(w) : vw \in E\}$. If $E$ is shadow-closed, then it can be seen that for each leaf $v$, $up(v)$ is an ancestor of $v$.

We use the standard notion of *contracting* an edge or a subset of nodes, see [6], [4, Chapter 1]. When we contract a subtree $T'$, then the (new) contracted node is denoted by the least common ancestor of the nodes in $T'$. For a set of links $F' \subseteq E$, we use $T/F'$ to denote the tree obtained by contracting each of the 2-edge connected components of $T + F' = (V, \widehat{E}_T \cup F')$ to a single node.

For a vector $x \in \mathbb{R}^E$ and any subset $J$ of $E$, $x(J)$ denotes $\sum_{e \in J} x_e$. Given several vectors $v^1, v^2, \ldots$, we write one of their convex combinations as $\sum_{i \in Z} \lambda_i v^i$; thus, $Z$ is a set of indices, and we have $\lambda_i \geq 0, \forall i \in Z$, and $\sum_{i \in Z} \lambda_i = 1$.

2.1. **Lift-and-project systems, and the Lasserre (Sum-of-squares) system.** We give a brief overview of the Lasserre system, [11]. In fact, our results can be stated and proved without going into the formalities of the Lasserre system. Essentially, we apply one well-known result about the Lasserre system, namely, the decomposition theorem of Rothvoß and Karlin-Mathieu-Nguyen, see [9, 12, 14]. The comprehensive recent survey by Rothvoß [14], presents this result and much more.

Let $A$ be an $m \times n$ matrix, and let $(LP_0) \quad \min\{c^T x \text{ s.t. } Ax \geq b\}$ be a linear programming relaxation of a binary integer programming problem; thus, each variable $x_1, \ldots, x_n$ is binary. We use the notation $A = (A_{\ell i})_{\ell \in [m], i \in [n]}$. (Below, we may use $(LP_0)$ to denote both the linear program above, and its feasible region; thus $LP_0 = \{x \in \mathbb{R}^n : Ax \geq b\}$.)

An $n \times n$ matrix $M$ is called positive semidefinite (p.s.d.) if $x^T M x \geq 0, \forall x \in \mathbb{R}^n$; $M \succcurlyeq 0$ denotes that $M$ is p.s.d.

For a positive integer $t$ and the ground set $[n] = \{1, \ldots, n\}$, let $\mathcal{P}_t$ denote the family of subsets of $[n]$ of size at most $t$, i.e., $\mathcal{P}_t = \{I : I \subseteq [n], |I| \leq t\}$; thus, each element of $\mathcal{P}_t$ is an "index set" of size $\leq t$. In this subsection, we use $I$ and $J$ to denote elements of $\mathcal{P}_t$ (whereas, in the rest of the paper, we use $J$ to denote an arbitrary subset of $E(G)$).

The $t$-th level of the Lasserre system (or, hierarchy), denoted $\text{LAS}_t(LP_0)$ consists of the vectors $y \in \mathbb{R}^{2^{[n]}}$ that satisfy

$$M_{t+1}(y) \equiv (y_{I \cup J})_{I,J \in \mathcal{P}_{t+1}} \succcurlyeq 0; \quad \hat{M}_t^\ell(y) \succcurlyeq 0, \ \forall \ell \in [m]; \quad y_\emptyset = 1,$$

where $\hat{M}_t^\ell(y) \equiv \left( \sum_{i=1}^n (A_{\ell i})(y_{I \cup J \cup \{i\}}) - (b_\ell)(y_{I \cup J}) \right)_{I,J \in \mathcal{P}_t}$.

In other words, the $t$-th level *moment matrix* of $y$, $M_t(y)$, is required to be p.s.d., and moreover, for each of the constraints of $(LP_0)$, namely, $\text{row}_\ell(A)x - b_\ell \geq 0$, $\ell \in [m]$, the associated $t$-th level *moment matrix of slacks*, $\hat{M}_t^\ell(y)$, is required to be p.s.d.

The index $t$ of each relaxation in the sequence of tightened relaxations is known as the *level*; the level of the original relaxation is defined to be zero. For any $t = O(1)$, it is known that the relaxation at level $t$ can be solved to optimality (up to a "small enough" additive error term) in polynomial time, assuming that the original relaxation has a polynomial-time separation oracle; additional mild conditions may be needed. Moreover, the relaxation at level $n$ is exact.

## 3. Starting LP relaxation

We formulate the "starting" LP relaxation $(LP_0)$ for TAP in this section. We apply the Lasserre (Sum-of-squares) system to $(LP_0)$. Our results are stated for the level $t$ tightening of $(LP_0)$ by the Lasserre system (where $t = O(1)$).

We say that a link $u_1 v_1$ is *overlapped* by another link $u_2 v_2$ if the following conditions hold
- either $u_1$ or $v_1$ belongs to $V(P_{u_2 v_2})$, and
- $P_{u_1 v_1}, P_{u_2 v_2}$ have one or more tree-edges in common.

We call a pair of links an *overlapping pair* if one link of the pair is overlapped by the other link. We call a subset $J$ of the links an *overlapping clique* if every pair of links in $J$ is an overlapping pair. The notion of overlapping pairs and the next lemma are critical for our starting LP relaxation of TAP.

**Lemma 3.1.** *For any shadow-closed instance of TAP, there exists an optimal solution $F \subseteq E$ that contains no overlapping pair.*

*Proof.* Suppose that $F' \subseteq E$ is an optimal solution that contains an overlapping pair $u_1 v_1, u_2 v_2$, and w.l.o.g., suppose that $u_1 \in V(P_{u_2 v_2})$ and the tree edge $\hat{e} = u_0 v_0$ is in both $P_{u_1 v_1}, P_{u_2 v_2}$. Then, there is a maximal (nonempty) prefix of (the edge sequence of) $P_{u_1 v_1}$ that is contained in $P_{u_2 v_2}$; let us denote this prefix by $P_{u_1, u_*}$ (clearly, $\hat{e} = u_0 v_0$ is in $P_{u_1, u_*}$). Then we could replace $u_1 v_1$ in $F$ by $u_* v_1$ (by the shadow-closed property of $E$) to get a new optimal solution that has fewer overlapping pairs.

The statement of the lemma follows by induction. $\square$

Based on this, we have the following LP relaxation of TAP:

$$(\boldsymbol{LP_0}) \quad \text{minimize} : \sum_{uv \in E} x_{uv}$$

$$\text{subject to} : \sum_{uv \in \hat{\delta}_E(\hat{e})} x_{uv} \geq 1 \quad \forall \hat{e} \in \widehat{E}_T$$

$$x_{u_1 v_1} + x_{u_2 v_2} \leq 1 \quad \forall \text{ overlapping pairs } \ u_1 v_1, u_2 v_2 \in E$$

$$0 \leq x \leq 1$$

Denote this LP relaxation of TAP by $(LP_0)$. Note that it is a tightening of the obvious "covering" LP relaxation of TAP (namely, $\min\{\sum_{e \in E} x_e \ : \ x(\hat{\delta}_E(\hat{e})) \geq 1, \forall \hat{e} \in \widehat{E}_T, \ x \geq 0\}$), because $(LP_0)$ has additional constraints for the overlapping pairs; these additional constraints are valid because $E$ is assumed to be shadow-closed.

## 4. Lasserre tightening and its properties

Let $\text{LAS}_t(LP_0)$ be the level $t$ tightening of $(LP_0)$ by the Lasserre system. Let $x \in \mathbb{R}^E$ be a feasible solution of $(LP_0)$. Define $\text{ones}(x) = \{uv : x_{uv} = 1\}$; thus, $\text{ones}(x)$ is the set of links of $x$-value one. Rothvoß, see [12, Theorem 2], formulated the following decomposition theorem for the Lasserre system, based on an earlier decomposition theorem due to Karlin-Mathieu-Nguyen [9].

**Theorem 4.1.** Let $J \subseteq E$. Let $k$ be a positive integer such that $|\text{ones}(x) \cap J| \leq k$ for every feasible solution $x$ of $(LP_0)$. Then for every feasible solution $y \in \text{LAS}_t(LP_0)$, where $t \geq k+1$, $y$ can be written as a convex combination: $y = \sum_{i \in Z} \lambda_i x^i$ such that $x^i$ is in $\text{LAS}_{t-k}(LP_0)$ and $x^i|_J$ is integral (i.e., $x^i_{uv}$ is integral for each $uv \in J$), for all $i \in Z$.

**Lemma 4.2.** Let $J \subseteq E$ be an overlapping clique. For every feasible solution $x$ of $(LP_0)$, we have $|\text{ones}(x) \cap J| \leq 1$.

*Proof.* Otherwise, there exist at least two distinct links $u_1v_1, u_2v_2 \in \text{ones}(x) \cap J$. Thus, $u_1v_1, u_2v_2$ is an overlapping pair but $x_{u_1v_1} = x_{u_2v_2} = 1$. This contradicts the overlapping constraints in $(LP_0)$. □

**Lemma 4.3.** Let $J \subseteq E$ be an overlapping clique. For every level $t \geq 2$ and every feasible solution $y$ of $\text{LAS}_t(LP_0)$, we have $y(J) \leq 1$.

*Proof.* By Lemma 4.2, $|\text{ones}(x) \cap J| \leq 1$ for any feasible solution $x$ of $(LP_0)$. Theorem 4.1 implies that $y$ can be written as a convex combination: $y = \sum_{i \in Z} \lambda_i x^i$ such that $x^i$ is in $\text{LAS}_1(LP_0)$ and $x^i|_J$ is integral for each $i \in Z$. Since $|\text{ones}(x) \cap J| \leq 1$ for every feasible solution $x$ of $(LP_0)$, we have $|\text{ones}(x^i) \cap J| \leq 1$. Since $x^i|_J$ is integral, $x^i(J) \leq 1$. This implies $y(J) \leq 1$. □

**Lemma 4.4.** Let $v$ be a leaf of $T$. Let $\hat{e}$ be a tree-edge in $\tilde{P}_v$. Then, $\hat{\delta}_E(\hat{e})$ (the set of links covering $\hat{e}$) is an overlapping clique. In particular, $\delta_E(v)$ is an overlapping clique.

*Proof.* Consider any two links $u_1v_1, u_2v_2$ in $\hat{\delta}_E(\hat{e})$. Let $q$ be the end-node of $\hat{e}$ with larger height. Each of the links $u_1v_1, u_2v_2$ must have an end-node in $P_{v,q}$, since each internal nodes of $\tilde{P}_v$ has exactly one child. Suppose that $v_1, v_2$ are the end-nodes in $P_{v,q}$, and w.l.o.g., assume that $h(v_1) \leq h(v_2)$. Then, observe that $u_1v_1, u_2v_2$ is an overlapping pair, because $v_1$ is in $P_{u_2v_2}$ and the tree-edge $\hat{e}$ is in both $P_{u_1v_1}$ and $P_{u_2v_2}$. Hence, $\hat{\delta}_E(\hat{e})$ is an overlapping clique.

For the second part, let $\hat{e}$ denote the unique tree-edge incident to the leaf $v$, and note that $\delta_E(v) = \hat{\delta}_E(\hat{e})$; clearly, $\hat{e} \in \tilde{P}_v$. □

**Corollary 4.5.** Let $t \geq 2$, and let $y \in \text{LAS}_t(LP_0)$ be a feasible solution to the $t$-th level of the Lasserre system. Let $v$ be a leaf of $T$. Let $\hat{e}$ be a tree-edge in $\tilde{P}_v$. Then, we have $y(\hat{\delta}_E(\hat{e})) = 1$. Moreover, we have $y(\delta_E(v)) = 1$.

*Proof.* Lemmas 4.3, 4.4 and the constraints of $(LP_0)$ imply that $y(\hat{\delta}_E(\hat{e})) = 1$.

For the last part, let $\hat{e} \in \tilde{P}_v$ denote the unique tree-edge incident to the leaf $v$, and note that $\delta_E(v) = \hat{\delta}_E(\hat{e}) = 1$. □

Recall that the matching polytope of the subgraph induced by the leaves, $G(L) = (L, E(L))$ is given by the following constraints:

$$x(\delta_{E(L)}(v)) \leq 1 \qquad \forall v \in L$$
$$x(E(U)) \leq \frac{|U|-1}{2} \qquad \forall U \subseteq L, |U| \text{ odd}$$
$$x \geq 0$$

Similarly, we can write the constraints for the matching polytope of the subgraph $G_{ntw} = (L, E_{ntw}) = (L, E(L) - E_{twin})$; this is the subgraph obtained from $G(L)$ by deleting all twin links.

The next result is essentially the result on the matching polytope from the survey of Rothvoß, see [14, Lemma 13,Sec 3.3], translated to our setting.

**Lemma 4.6.** *Let $\epsilon > 0$, and let $y \in \text{LAS}_{\frac{1}{2\epsilon}+1}(LP_0)$ be a feasible solution to the $(1 + \frac{1}{2\epsilon})$-th level of the Lasserre system. Then, $\frac{y|_{E(L)}}{1+\epsilon}$ is in the matching polytope of $G(L) = (L, E(L))$. Similarly, $\frac{y|_{E_{ntw}}}{1+\epsilon}$ is in the matching polytope of $G_{ntw} = (L, E_{ntw})$.*

*Proof.* By Corollary 4.5, we have $y(\delta_{E(L)}(v)) \leq 1, \forall v \in L$.

Thus, it suffices to prove that $y|_{E(L)}(E(U)) \leq \frac{|U|-1}{2}(1+\epsilon)$ for all subsets $U$ of $L$ of odd size.

First, consider odd sets $U \subseteq L$ with $|U| > \frac{1}{\epsilon} + 1$. Clearly, we have $y|_{E(L)}(E(U)) \leq \frac{|U|}{2}$ since $y(\delta_{E(L)}(v)) \leq 1, \forall v \in L$. Also, observe that $\frac{|U|}{2} = \frac{|U|-1}{2}(1+\frac{1}{|U|-1}) < \frac{|U|-1}{2}(1+\frac{1}{\frac{1}{\epsilon}+1-1}) = \frac{|U|-1}{2}(1+\epsilon)$. Hence, $y|_{E(L)}(E(U)) \leq \frac{|U|-1}{2}(1+\epsilon)$.

Now, consider odd sets $U \subseteq L$ with $|U| \leq \frac{1}{\epsilon} + 1$. We apply the decomposition theorem, Theorem 4.1. Note that for any feasible solution $x$ of $(LP_0)$, by Lemmas 4.2, 4.4, we have $|ones(x) \cap E(U)| \leq \frac{|U|-1}{2} \leq \frac{1}{2\epsilon}$. Since $y$ is a feasible solution to the $(1 + \frac{1}{2\epsilon})$-th level of the Lasserre system, $y$ can be written as a convex combination: $y = \sum_{i \in Z} \lambda_i x^i$ such that $x^i$ is in $\text{LAS}_1(LP_0)$ and $x^i|_{E(U)}$ is integral. Hence, for each $i \in Z$, $x^i(E(U)) \leq \frac{|U|-1}{2}$. Consequently, $y|_{E(L)}(E(U)) \leq \frac{|U|-1}{2}$.
This completes the proof. □

**Lemma 4.7.** *Let $t \geq 4$, and let $y \in \text{LAS}_t(LP_0)$ be a feasible solution to the $t$-th level of the Lasserre system. Let $s$ be a stem, and let $vw$ be the twin link of $s$. Then we have $y(vw) \leq y(\delta_E(s))$.*

*Proof.* Let $J = \delta_E(v) \cup \delta_E(w)$. Note that both $\delta_E(v)$ and $\delta_E(w)$ are overlapping cliques, by Lemma 4.4. For any feasible solution $x$ of $(LP_0)$, by Lemma 4.2, $|ones(x) \cap \delta_E(v)| \leq 1$ and $|ones(x) \cap \delta_E(w)| \leq 1$. This implies $|ones(x) \cap J| \leq 2$. Hence, by Theorem 4.1, $y$ can be written as a convex combination $\sum_{i \in Z} \lambda_i x^i$ such that $x^i \in \text{LAS}_2(LP_0)$ and $x^i|_J$ is integral, $\forall i \in Z$. Since $vw \in J$, we have $x^i(vw)$ is integral. Hence, either $x^i(vw) = 0$ or $x^i(vw) = 1$. Let $Z_1 = \{i : x^i(vw) = 1\}$. Then $y(vw) = \sum_{i \in Z_1} \lambda_i x^i(vw)$.

Consider $x^i$ for $i \in Z_1$. Clearly, the link $vw$ covers each tree-edge $\hat{e} \in \tilde{P}_v \cup \tilde{P}_w$. Since $x^i \in \text{LAS}_2(LP_0)$, by Corollary 4.5, we have $x^i(\hat{\delta}_E(\hat{e})) = 1$ for each tree-edge $\hat{e} \in \tilde{P}_v \cup \tilde{P}_w$. Hence, for each tree-edge $\hat{e} \in \tilde{P}_v \cup \tilde{P}_w$, we see that $\hat{\delta}_E(\hat{e})$ has a unique link with positive $x^i$-value, namely, $vw$.

Now, consider the tree-edge $\hat{e}_s$ from $s$ to its parent. By the constraints of $(LP_0)$, we have $x^i(\hat{\delta}_E(\hat{e}_s)) \geq 1$. Notice that every link in $\hat{\delta}_E(\hat{e}_s)$ with positive $x^i$-value must have $s$ as its end-node in $T_s$; otherwise, if such a link has an end-node at some other node of $T_s$, then one of the tree-edges $\hat{e} \in T_s = \tilde{P}_v \cup \tilde{P}_w$ will have $x^i(\hat{\delta}_E(\hat{e})) > 1$, thus contradicting the equation given by Corollary 4.5. Therefore, $x^i(\delta_E(s)) \geq 1 = x^i(vw), \forall i \in Z_1$. Consequently, $y(vw) = \sum_{i \in Z_1} \lambda_i x^i(vw) \leq \sum_{i \in Z_1} \lambda_i x^i(\delta_E(s)) \leq \sum_{i \in Z} \lambda_i x^i(\delta_E(s)) = y(\delta_E(s))$. This completes the proof. □

## 5. Lower bound

The next lemma leads to the lower-bounding function employed by our analysis.

**Lemma 5.1.** *For any feasible solution $x \in \mathbb{R}^E$ of $(LP_0)$, we have*

$$x(E) \geq \frac{2}{3}\sum_{v \in L} x(\delta_E(v)) + \frac{1}{3}\sum_{v \in \mathcal{S}} x(\delta_E(v)) + \frac{1}{3}\sum_{v \in \mathcal{R}} x(\delta_E(v)) - \frac{1}{3}x(E(L)).$$

*Proof.*

$$\begin{aligned}
x(E) &= \frac{1}{2}\sum_{v\in L}x(\delta_E(v)) + \frac{1}{2}\sum_{v\in S}x(\delta_E(v)) + \frac{1}{2}\sum_{v\in R}x(\delta_E(v)) \\
&\geq \frac{1}{2}\sum_{v\in L}x(\delta_E(v)) + \frac{1}{3}\sum_{v\in S}x(\delta_E(v)) + \frac{1}{3}\sum_{v\in R}x(\delta_E(v)) + \frac{1}{6}(x(E(L,S)) + x(E(L,R))) \\
&= \frac{1}{2}\sum_{v\in L}x(\delta_E(v)) + \frac{1}{3}\sum_{v\in S}x(\delta_E(v)) + \frac{1}{3}\sum_{v\in R}x(\delta_E(v)) + \frac{1}{6}\sum_{v\in L}x(\delta_E(v)) - \frac{1}{3}x(E(L)) \\
&= \frac{2}{3}\sum_{v\in L}x(\delta_E(v)) + \frac{1}{3}\sum_{v\in S}x(\delta_E(v)) + \frac{1}{3}\sum_{v\in R}x(\delta_E(v)) - \frac{1}{3}x(E(L)).
\end{aligned}$$

□

Let $M$ denote a maximum matching of $G_{ntw} = (L, E_{ntw}) = (L, E(L)-E_{twin})$; thus, $M$ is a maximum matching of the leaf-to-leaf "non twin links".

**Lemma 5.2.** *Let $\epsilon > 0$ be a constant, and let $t = \max\{\frac{1}{2\epsilon}+1, 4\}$. Let $y^* \in \mathbb{R}^E$ be a feasible solution of the t-th level of the Lasserre system, $\text{LAS}_t(LP_0)$. Then,*

$$y^*(E) \geq \frac{2}{3}|L| + \frac{1}{3}\sum_{v\in R}y^*(\delta_E(v)) - \frac{1}{3}(1+\epsilon)|M|.$$

*Proof.* We will prove that

$$(1+\epsilon)|M| \geq y^*(E(L)) - \sum_{v\in S}y^*(\delta_E(v)).$$

The result then follows from this inequality and Lemma 5.1, by using Corollary 4.5 which implies that $|L| = \sum_{v\in L}y^*(\delta_E(v))$. To see this, apply Lemma 5.1 to $y^*$ (and replace $\sum_{v\in L}y^*(\delta_E(v))$ by $|L|$) to get

$$y^*(E) \geq \frac{2}{3}|L| + \frac{1}{3}\sum_{v\in R}y^*(\delta_E(v)) - \frac{1}{3}\left(y^*(E(L)) - \sum_{v\in S}y^*(\delta_E(v))\right) \geq \frac{2}{3}|L| + \frac{1}{3}\sum_{v\in R}y^*(\delta_E(v)) - \frac{1}{3}(1+\epsilon)|M|.$$

Consider the inequality $(1+\epsilon)|M| \geq y^*(E(L)) - \sum_{v\in S}y^*(\delta_E(v))$.

By Lemma 4.6, there exist matchings $M_1, M_2, \ldots, M_k$ on $G_{ntw} = (L, E_{ntw})$ such that $\frac{y^*|_{E_{ntw}}}{1+\epsilon}$ can be written as a convex combination of incidence vectors of these matchings, i.e., $y^*|_{E_{ntw}} = (1+\epsilon)\sum_{i\in Z}\lambda_i x^i$ where $x^i$ is the incidence vector of $M_i$. Hence, $y^*(E(L)) = y^*(E_{twin}) + y^*(E_{ntw}) = y^*(E_{twin}) + (1+\epsilon)\sum_{i\in Z}\lambda_i x^i(E_{ntw})$. Observe that $M$ is a maximum matching of $G_{ntw} = (L, E_{ntw})$, therefore, $(1+\epsilon)\sum_{i\in Z}\lambda_i x^i(E_{ntw}) \leq (1+\epsilon)|M|$. Moreover, by Lemma 4.7, we have $y^*(E_{twin}) \leq \sum_{v\in S}y^*(\delta_E(v))$. Thus, we have $y^*(E(L)) \leq \sum_{v\in S}y^*(\delta_E(v)) + (1+\epsilon)|M|$, that is, $(1+\epsilon)|M| \geq y^*(E(L)) - \sum_{v\in S}y^*(\delta_E(v))$. This completes the proof. □

Let $L_{unm}(T)$ denote the set of $M$-exposed leaf nodes, that is, the set of leaves that are not covered by $M$; we may abbreviate this notation to $L_{unm}$, if there is no danger of confusion. Observe that $|L_{unm}| = |L| - 2|M|$. By Lemma 5.2, we have

$$y^*(E) \geq \frac{2}{3}|L_{unm}| + (1-\frac{1}{3}\epsilon)|M| + \frac{1}{3}\sum_{v\in R}y^*(\delta_E(v)).$$

Multiplying both sides by $\frac{9}{5}$, we get

$$\frac{9}{5}y^*(E) \geq \frac{6}{5}|L_{unm}| + (\frac{9}{5} - \frac{3}{5}\epsilon)|M| + \frac{3}{5}\sum_{v \in \mathcal{R}} y^*(\delta_E(v)).$$

Note that $|M| \leq \frac{1}{2}|L| = \frac{1}{2}\sum_{v \in L} y^*(\delta_E(v)) \leq y^*(E)$. Hence, we have

$$(\frac{9}{5} + \epsilon)y^*(E) \geq (\frac{9}{5} + \frac{3}{5}\epsilon)y^*(E) \geq \frac{6}{5}|L_{unm}| + \frac{9}{5}|M| + \frac{3}{5}\sum_{v \in \mathcal{R}} y^*(\delta_E(v)).$$

Let $M' \subseteq M$ be the set consisting of links $e$ such that there exists a stem $s$ with leaves $u_s, v_s$ (in $T_s$) such that one of $u_s, v_s$, say $u_s$, is an end-node of $e$ and the other leaf $v_s$ is $M$-exposed. Thus, $M' = \{e \in M \,:\, \exists s \in \mathcal{S} \,:\, M \cap \delta_E(V(T_s)) = \{e\}\}$. Clearly, each edge in $M'$ is paired with an $M$-exposed leaf of a stem, hence, $|L_{unm}| \geq |M'|$.

In the last inequality (above), we replace the term $\frac{6}{5}|L_{unm}| + \frac{9}{5}(|M-M'| + |M'|)$ by $|L_{unm}| + \frac{9}{5}|M-M'| + 2|M'|$ to get

$$(\frac{9}{5} + \epsilon)y^*(E) \geq |L_{unm}| + \frac{9}{5}|M-M'| + 2|M'| + \frac{3}{5}\sum_{v \in \mathcal{R}} y^*(\delta_E(v)),$$

and we "relax" this to get

$$(\frac{9}{5} + \epsilon)y^*(E) \geq |L_{unm}| + \frac{3}{2}|M-M'| + 2|M'| + \frac{1}{2}\sum_{v \in \mathcal{R}} y^*(\delta_E(v)). \quad \text{(LB)}$$

## 6. Algorithm, coupons and tickets

This section presents the approximation algorithm and its analysis. We state our main result for (unweighted) TAP:

**Theorem 6.1.** *Let $\epsilon > 0$ be a constant, and let $t = \max\{4, \frac{1}{2\epsilon} + 1\}$. The integrality ratio of $\text{LAS}_t(LP_0)$ is $\leq \frac{9}{5} + \epsilon$; moreover, there is a polynomial-time algorithm for finding a feasible solution of size $\leq (\frac{9}{5} + \epsilon)y^*(E)$, where $y^*$ is an optimal solution of $\text{LAS}_t(LP_0)$.*

Let $y^* \in \mathbb{R}^E$ be an optimal solution of the $t$-th level of the Lasserre system, $\text{LAS}_t(LP_0)$, where $t = \max\{4, 1 + \frac{1}{2\epsilon}\}$, where $\epsilon > 0$ is a constant. Our goal is to show that our algorithm finds a feasible solution (a set of links that covers $T$) whose size (i.e., number of links) is $\leq (\frac{9}{5} + \epsilon)y^*(E)$. To achieve the goal, we use the right-hand side (r.h.s.) of the inequality (LB) (at the end of Section 5) as our "potential function".

As discussed in Section 1, our algorithm and analysis essentially follow the algorithm and analysis of [6, Sections 3,4], although, there are several important differences.

We use $I$ to denote the set of links (currently) picked by our algorithm; initially, $I$ is the empty set, and at termination, $I$ covers $T$, i.e., $T + I = (V, E(T) \cup I)$ is 2-edge connected. Recall that $T/I$ denotes the tree obtained by contracting each of the 2-edge connected components of $T + I$ to a single node.

Each of the contracted nodes of the current tree $T/I$ is called a *compound node*, see [6, Section 3.2]; thus, each compound node corresponds to a set of two or more nodes of $V(T)$. Each of the other nodes of $T/I$ is called an *original node*.

Following [6], our algorithm maintains the following assignment of *coupons* and *tickets* to the nodes of $T/I$ and the links of $M/I$ (we start with an assignment to the nodes of $T$ and the links of $M$, since $I$ is empty at the start). The motivation is to distribute the "credit" from our "potential function" (namely, $|L_{unm}(T)| + \frac{3}{2}|M-M'| + 2|M'| + \frac{1}{2}\sum_{v \in \mathcal{R}} y^*(\delta_E(v))$) to these nodes and links, such that the "cost" of adding new links to $I$ and updating coupons can be "paid" from the coupons and tickets "associated with" the new links:

- every unmatched original leaf has one coupon,
- every compound node has one coupon,
- every link in $M-M'$ has $\frac{3}{2}$ coupons,
- every link in $M'$ has 2 coupons, and
- the root $r$ has one coupon.

Moreover, we assign tickets (these are fractional values) as follows.

- every node $v \in \mathcal{R}$ is assigned $\frac{1}{2} y^*(\delta_E(v))$ tickets.

Here is an outline of the algorithm; also see [6, Section 3.4]. We start with $T' := T$ and $I := \emptyset$. Then we repeatedly find a set of links $I' \subseteq E-I$ such that the addition of $I'$ to $I$ results in a single new compound node, and moreover, the number of coupons and tickets associated with $I'$ is $\geq |I'| + 1$ (the "+1" is needed for the coupon of the new compound node). This step is repeated until $T'$ is a single node, that is, until $T + I$ is 2-edge connected.

For a set of nodes $U \subseteq V(T/I)$, we define the *potential* $\Phi(U)$ to be the sum of the coupons/tickets of the nodes in $U$, thus, $\Phi(U) = \frac{1}{2} \sum_{v \in U \cap \mathcal{R}} y^*(\delta_E(v)) +$ (number of coupons of nodes in $U$).

For a set of links $J \subseteq E-I$, we define $J^*$ to be the set of all links $vq \in E-I$ such that for each tree-edge $\hat{e}$ of $P_{v,q}$ there exists a link in $J$ that covers $\hat{e}$ (i.e., $\exists uw \in J : P_{u,w} \ni \hat{e}$); thus, $J^* \subseteq E-I$ is the set of links that have both end-nodes in a subtree of $T/I$ that is "covered" by $J$. For a set of links $J \subseteq E-I$, we define the potential $\Phi(J)$ to be the sum of the number of coupons of $J^*$ and the number of coupons/tickets of the nodes of $T/I$ covered by $J$; thus, $\Phi(J) = 2|J^* \cap M'| + \frac{3}{2}|J^* \cap (M - M')| + \Phi(\{v : \exists uw \in J : P_{u,w} \ni v\})$.

We call a link $e = uw \in E-I$ a *good link* if its potential $\Phi(uw)$ is $\geq 2$. Similarly, we call a set of links $J \subseteq E-I$ a *good link set* if the contraction of all links in $J$ results in a single (new) compound node, and moreover, $\Phi(J) \geq |J| + 1$.

We mention that each coupon/ticket is used at most once by the algorithm/analysis. In other words, the "cost" of the set of links $I'$ contracted by a particular iteration (namely, $|I'| + 1$) is paid by the coupons/tickets associated with $I'$; but, these specific coupons/tickets are never used again in the execution of the algorithm; of course, the algorithm creates one new coupon for the new compound node, and that new coupon could be used later in the execution, however, the algorithm has already paid for the new coupon from the coupons/tickets associated with $I'$. (Also see the discussion under "Tickets" in [6, Section 3.2].)

6.1. **Pre-processing for stems.** Before the main algorithm is applied, we apply a pre-processing step that "eliminates" all the stem nodes by contracting them into compound nodes. Also see [6, Section 3.3], in particular, "greedy contractions", Invariant 3.6, and Lemma 3.8.

The stems can be processed in any order, although the processing of one stem may result in contractions that create a new compound node that "contains" several stems. Consider any stem $s$ and let $u_s, w_s$ denote the two leaves in $T_s$.

- If both $u_s$ and $w_s$ are $M$-exposed, then we pick the twin link $u_s w_s$ by letting $I' := \{u_s w_s\}$. Observe that each of the leaves $u_s, w_s$ has one coupon.
- If exactly one of $u_s$ and $w_s$ is $M$-exposed, say, $u_s$ is $M$-exposed, then $w_s$ is incident to a link in $M' \subseteq M$, say $w_s q_s \in M'$; observe that $w_s q_s$ has two coupons, and the leaf $u_s$ has one coupon. We pick both the links $w_s q_s$ and $u_s w_s$ (twin link) by letting $I' = \{w_s q_s, u_s w_s\}$.
- If neither $u_s$ nor $w_s$ is $M$-exposed, then $u_s$ and $w_s$ are incident to two different links of $M$, say $u_s p_s \in M$ and $w_s q_s \in M$ (recall that $M$ contains no twin links); observe that each of these links has $\frac{3}{2}$ coupons. We pick both the links $u_s p_s$ and $w_s q_s$ by letting $I' = \{u_s p_s, w_s q_s\}$.

In each of these three cases, the algorithm creates one new compound node, by adding a set of links $I'$ (where $|I'| = 1$ or $|I'| = 2$) to the current set $I$, and by contracting $I'$ to obtain $T/(I \cup I')$. (At the start, $I = \emptyset$.) We assign one coupon to the new compound node.

It is easily seen that for each case in the pre-processing step, the set of contracted links $I'$ has $\Phi(I') \geq |I'| + 1$, thus $I'$ is a good link set. The following lemma summarizes the effect of the pre-preprocessing.

**Lemma 6.2.** *After the pre-preprocessing step, all the stems are contracted into compound nodes. Moreover, all the links in $M'$ are contracted.*

**6.2. Simple contractions.** We define a *simple contraction* to be one of the following types of contractions; each of these contractions applies to a single good link. (We mention that our simple contractions are special cases of the "greedy 1-contractions" of [6, Section 3.3].)

- For the current tree, consider a leaf-to-leaf link $uw$ such that each end-node owns a coupon; thus each of $u, w$ is either a compound node that is a leaf, or an original leaf node that is $M$-exposed. Observe that $uw$ is a good link because $\Phi(uw) \geq 2$.
- For the current tree, consider a link $uw \in M$ such that $P_{u,w}$ contains at least one compound node. Again, note that $uw$ is a good link because $\Phi(uw) \geq 2$.

**6.3. Semiclosed trees and deficient 3-leaf configurations.** The notion of semiclosed trees is due to Even, et al., based on earlier work by Nagamochi [7]; also, see [6, Definition 2.3].

A rooted subtree $T_u$ of $T$ is called *semiclosed* w.r.t. (with respect to) a matching $\hat{M} \subseteq E$ if the following conditions hold:

(i) $\hat{M} \cap \delta_E(V(T_u)) = \emptyset$, that is, each link in $\hat{M}$ either has both end-nodes in $T_u$ or has no end-node in $T_u$.
(ii) Every link incident to an $\hat{M}$-exposed leaf of $T_u$ has both end-nodes in $T_u$; in other words, every link with exactly one end-node in $T_u$ has that end-node either at a non-leaf node or at an $\hat{M}$-covered leaf node.

For any rooted subtree $T_u$ of the algorithm's current tree $T' = T/I$, let $C(T_u)$ denote the set of non-leaf (i.e., internal) compound nodes in $T_u$. For any rooted subtree $T_u$ of $T'$ and the matching $M$ (recall that $M$ is a maximum matching of the leaf-to-leaf "non twin links") we use $M(T_u)$ to denote the set of links in $M$ that have both end-nodes in $T_u$, and we use $L_{unm}(T_u)$ to denote the set of $M$-exposed leaves of $T_u$; moreover, we use $\mathcal{R}(T_u)$ to denote the set of nodes $V(T_u) \cap \mathcal{R}$.

For any semiclosed tree $T_u$ w.r.t. $M$, we define its "basic cover" $B(T_u)$ as follows:

$$B(T_u) \equiv M(T_u) \cup \{up(w)w \; : \; w \in L_{unm}(T_u)\}.$$

By a *minimally semiclosed tree* $T_u$ we mean that $T_u$ is semiclosed but none of the proper rooted subtrees of $T_u$ is semiclosed.

**Lemma 6.3** (Even, et al., [6]). *Let $T_u$ be a minimally semiclosed tree. Then $B(T_u)$ covers all the tree-edges of $T_u$ (i.e., $T_u + B(T_u)$ is 2-edge connected).*

For a rooted tree $T_u$ of the algorithm's current tree $T' = T/I$, we define

$$\text{credit}(T_u) \equiv \frac{3}{2}|M(T_u)| + \Phi(V(T_u)) \;=\; \frac{3}{2}|M(T_u)| + |L_{unm}(T_u)| + |C(T_u)| + \frac{1}{2}\sum_{v \in \mathcal{R}(T_u)} y^*(\delta_E(v))$$

Observe that the leaves in $L_{unm}(T_u)$ may be either original nodes or compound nodes, but, in either case, each such node has one coupon. Moreover, each non-leaf compound node of $T_u$ has one coupon.

We call a semiclosed tree $T_u$ *good* if $credit(T_u) \geq |B(T_u)| + 1$. Clearly, if a minimally semiclosed tree $T_u$ is good, then we may add $B(T_u)$ to the set of picked links $I$ and contract $T_u$ to a new compound node (with one coupon), and we can "pay" for this from $credit(T_u)$.

Even, et al., introduced the notion of "deficient trees", see [6, Definition 4.7] and Figure 1 of [6]; these are two specific "configurations" that cannot be handled by their "usual methods"; each of these configurations consists of a rooted subtree with three leafs and some incident links. We will call these *deficient 3-leaf configurations*, and following [6] we define them as follows:

Suppose that a rooted subtree $T_v$ of $T/I$ has exactly three leaves $a, b_1, b_2$, no non-leaf compound node, $M(T_v)$ has one link $b_1b_2$, and $T_v$ is not leaf-closed (a link incident to a leaf of $T_v$ has one end-node that is not in $T_v$). Moreover, suppose that $b_1, b_2$ can be ordered such that (i) $ab_1$ is a link and the contraction of $ab_1$ does not create a new leaf, and (ii) $b_2w$ is a link such that $w$ is not in $T_v$. (Moreover, if both orderings of $\{b_1, b_2\}$ satisfy (i), (ii), then assume that $up(b_2)$ is an ancestor of $up(b_1)$.) Then we call $T_v$ a deficient 3-leaf configuration.

The last part of [6] has a procedure that is applied when all minimally semiclosed trees satisfy the properties of a deficient 3-leaf configuration; this procedure finds another semiclosed tree $\hat{T}_u$ that is good as well as a cover for this tree of size $|B(\hat{T}_u)|$, and after that, the algorithm continues by applying their "usual methods".

**Remark.** The procedure in [6] can be easily extended to handle a small generalization: any of the tree-edges of a deficient 3-leaf configuration may be subdivided, i.e., any of the tree-edges may be replaced by a path of $\geq 1$ tree-edges (the internal nodes of these paths have degree 2 in the tree). In what follows, we use this (more general) notion of deficient 3-leaf configurations.

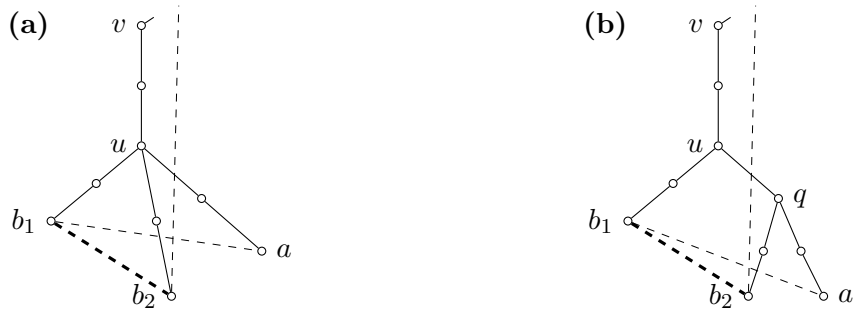

FIGURE 1. Illustration of deficient 3-leaf configurations.

6.4. **Main algorithm.** A sketch of the main algorithm follows. We start with $I := \emptyset$ ($I$ is the set of links picked by the algorithm) and $T' := T$ ($T'$ is the current tree $T/I$).

```
while T' is not a single node do
    apply all simple contractions ;
    if there exists a minimally semiclosed tree T_u that is good then
        | add B(T_u) to I, contract T_u to a new compound node, update T'
    else
        every minimally semiclosed tree must satisfy the properties of a deficient 3-leaf
        configuration; apply the procedure of [6, Section 4.3] to obtain a good semiclosed tree
        T̂_u; add a set of links B̂ of size |B(T̂_u)| to I, contract T̂_u to a new compound node,
        update T'
    end
end
```
**Algorithm 1:** Main Algorithm

The algorithm maintains the following invariant throughout.

**Invariant.** Suppose that no simple contractions are applicable. Then
  (1) Every compound node is $M$-exposed. In other words, both end-nodes of every link of $M$ are original leaf nodes.
  (2) There exist no links between $M$-exposed leaves.

6.5. **Analysis of the main algorithm.** We complete the analysis of the algorithm in this subsection. We prove that if there exist no minimal semiclosed trees that are good, then every semiclosed tree $T_v$ forms a deficient 3-leaf configuration (that is, the conditions of a deficient 3-leaf configuration apply to $T_v$). Our analysis is similar to that of [6, Section 4.2], but some new difficulties arise since our analysis is based on a fractional solution $y^*$ whereas the analysis of [6] is based on an integral solution. We are able to handle these difficulties by applying properties of solutions to the Lasserre system.

**Theorem 6.4.** *Let $T_v$ be a semiclosed tree that is not good. Then the conditions of a deficient 3-leaf configuration apply to $T_v$.*

*Proof.* Suppose that $T_v$ is not good; thus, $credit(T_v) < |B(T_v)|+1$. We have $credit(T_v) = \frac{3}{2}|M(T_v)| + |L_{unm}(T_v)| + |C(T_v)| + \frac{1}{2}\sum_{w\in\mathcal{R}(T_v)} y^*(\delta_E(w))$, and $|B(T_v)| = |M(T_v)| + |L_{unm}(T_v)|$. Let $ticket(T_v)$ denote $\frac{1}{2}\sum_{w\in\mathcal{R}(T_v)} y^*(\delta_E(w))$, the number of tickets assigned to nodes in $T_v$. Since $credit(T_v) < |B(T_v)|+1$, we must have $\frac{1}{2}|M(T_v)| + |C(T_v)| + ticket(T_v) < 1$.

Clearly, $v \neq r$, that is, $v$ is not the root; otherwise, since $r$ has one coupon (thus, it is the same as a non-leaf compound node), we would have $credit(T_v) \geq |B(T_v)| + 1$.

We must have $|M(T_v)| \leq 1$, $|C(T_v)| = 0$, and $ticket(T_v) < 1$. We consider several cases, depending on the sizes of $C(T_v), M(T_v)$ and $L_{unm}(T_v)$.

Moreover, we assume that every link with at least one end-node in $T_v$ has positive $y^*$ value; links with zero $y^*$ value are not relevant for our arguments, and we delete them.

**Case a.** $C(T_v) = \phi, M(T_v) = \phi$. Observe that $T_v$ has no links between any two of its leaves; this follows from simple contractions since all leaves of $T_v$ are $M$-exposed. Now, consider all links incident to leaves of $T_v$. Since $C(T_v) = \phi$, this will contribute at least $1/2$ to $ticket(T_v)$. Additionally, consider the links that cover the tree-edge between $v$ and its parent. Since $T_v$ is semiclosed, each such link has one of its end-nodes at a non-leaf node of $T_v$. This contributes $1/2$ to $ticket(T_v)$. To sum up, we have $ticket(T_v) \geq 1$, and thus we have a contradiction.

**Case b.** $C(T_v) = \phi, |M(T_v)| = 1, L_{unm}(T_v) = \phi$. Let $uw \in M(T_v)$, i.e., $uw$ is the link of $M$ with both end-nodes in $T_v$. There exists no compound node in $P_{u,w}$, by simple contractions. Then, it can be seen that $T_v$ contains a stem and $uw$ is a twin link. This is a contradiction.

**Case c.** $C(T_v) = \phi$, $|M(T_v)| = 1$, $|L_{unm}(T_v)| \geq 2$. Let $uw \in M(T_v)$, i.e., $uw$ is the link of $M$ with both end-nodes in $T_v$. All nodes in $P_{u,w}$ are original nodes, by simple contractions. By Corollary 4.5, we have $y^*(\delta_E(u)) = 1 = y^*(\delta_E(w))$. Now consider the links incident to the leaves in $L_{unm}(T_v)$, and the links covering the tree-edge between $v$ and its parent. All these links have an end-node in $T_v$, and the sum of their $y^*$ values is $\geq 3$. Moreover, by simple contractions and the fact that $T_v$ is semiclosed, the end-nodes of these links in $T_v$ must be in $\{u, w\} \cup \mathcal{R}(T_v)$. Since $y^*(\delta_E(u)) + y^*(\delta_E(w)) = 2$, the links incident to $\mathcal{R}(T_v)$ must have $y^*$ values summing to $\geq 1$. Hence, $ticket(T_v) \geq \frac{1}{2}$, and finally we have $ticket(T_v) + \frac{1}{2}|M(T_v)| \geq 1$. This is a contradiction.

**Case d.** $C(T_v) = \phi, |M(T_v)| = 1, |L_{unm}(T_v)| = 1$. We must have $ticket(T_v) < \frac{1}{2}$, since $T_v$ is not good. Thus, $T_v$ has exactly three leaves, where the two leaves incident to the link in $M(T_v)$ must be original leaf nodes, and the $M$-exposed leaf may be a compound node or an original leaf. We denote the $M$-exposed leaf by $a$, and the two $M$-covered leaves by $b_1, b_2$. Now, our goal is to show that $T_v$ satisfies all the conditions of a deficient 3-leaf configuration.

Observe that $T_v$ is isomorphic (up to nodes of degree 2) to one of two tree-configurations:
(1) $v$ has a descendant $u$ that has three children, and each of the leaves $a, b_1, b_2$ is a descendant of one of these children of $u$. (See Figure 1(a) and Figure 1(b) of [6].)
(2) $v$ has a descendant $u$ that has two children, and one of these children has a descendant $q$ that has two children, and each of the leaves $a, b_1, b_2$ is a descendant of one of the children of $q$ or of the child of $u$ that is not an ancestor of $q$. (See Figure 1(b) and Figure 1(a) of [6].) Moreover, it can be seen that $a$ is a descendant of $q$. W.l.o.g. assume that $b_2$ is the other "leaf descendant" of $q$.

**Case d.1** Suppose that $T_v$ is isomorphic (up to nodes of degree 2) to tree-configuration (1).

First, suppose that $y^*(ab_1) = 0$ and $y^*(ab_2) = 0$. Since $T_v$ is semiclosed, all links incident to $a$ have their other end-nodes in $\{b_1, b_2\} \cup \mathcal{R}(T_v)$. Thus, $ticket(T_v) \geq \frac{1}{2}$ due to the links incident to $a$, since $y^*(\delta_E(a)) \geq 1$.

Next, suppose that both $y^*(ab_1)$ and $y^*(ab_2)$ are positive. If either $b_1$ or $b_2$ is incident to a link that has its other end-node in $V - V(T_v)$, then it can be seen (by the symmetry of $b_1, b_2$) that $T_v$ satisfies the conditions for a deficient 3-leaf configuration (Figure 1(b) of [6]). Now, assume that neither $b_1$ nor $b_2$ is incident to any link that has its other end-node in $V - V(T_v)$. Consider the links covering the tree-edge between $v$ and its parent. The sum of the $y^*$ values of these links is $\geq 1$. Each of these links has an end-node in $\mathcal{R}(T_v)$ (none has an end-node in $\{b_1, b_2\}$). Thus, $ticket(T_v) \geq \frac{1}{2}$, and this is a contradiction.

Now, w.l.o.g. assume that $y^*(ab_1)$ is positive, and $y^*(ab_2) = 0$. If $b_2$ is incident to a link that has its other end-node in $V - V(T_v)$, then it can be seen that $T_v$ satisfies the conditions for a deficient 3-leaf configuration (Figure 1(b) of [6]). Otherwise, we get a contradiction as follows. Consider the links incident to $a$, and the links covering the tree-edge between $v$ and its parent. The sum of the $y^*$ values of these links is $\geq 2$. Each of these links has an end-node in $\{b_1\} \cup \mathcal{R}(T_v)$ (none has an end-node in $\{b_2\}$). By Corollary 4.5, $y^*(\delta_E(b_1)) = 1$. Hence, the sum of the $y^*$ values of the links incident to $\mathcal{R}(T_v)$ is $\geq 1$; thus, $ticket(T_v) \geq \frac{1}{2}$, and this is a contradiction.

**Case d.2** Suppose that $T_v$ is isomorphic (up to nodes of degree 2) to tree-configuration (2). Note that $b_1$ and $b_2$ (the end-node of the unique link in $M(T_v)$) are original leaf nodes. By Lemma 4.4, both $\delta_E(b_1)$ and $\delta_E(b_2)$ are overlapping cliques. Let $J = \delta_E(b_1) \cup \delta_E(b_2)$. For any feasible solution $x \in \mathbb{R}^E$ of $(LP_0)$, we have $|ones(x) \cap J| \leq 2$. Hence, by Theorem 4.1, and the fact that $t \geq 4$, $y^*$ can be written as a convex combination $\sum_{i \in Z} \lambda_i x^i$ such that $x^i \in \text{LAS}_{t-2}(LP_0)$ and $x^i$ is integral on $J$ for each $i \in Z$. If each $x^i$, $i \in Z$ has $\frac{1}{2} \sum_{w \in \mathcal{R}(T_v)} x^i(\delta_E(w)) \geq \frac{1}{2}$, then since $y^*$ is a convex combination of the $x^i$, we have $ticket(T_v) \geq \frac{1}{2}$. This gives a contradiction. Hence, there exists an $i_0 \in Z$ such that $\sum_{w \in \mathcal{R}(T_v)} x^{i_0}(\delta_E(w)) < 1$ (i.e., $ticket(T_v)$ w.r.t. $x^{i_0}$ is $< \frac{1}{2}$). One key observation is that $x^{i_0}$ is integral on $J$. Hence, either $x^{i_0}(b_1b_2) = 0$ or $x^{i_0}(b_1b_2) = 1$. We consider each of these cases.

In what follows, we assume that every link with at least one end-node in $T_v$ has positive $x^{i_0}$ value; links with zero $x^{i_0}$ value are not relevant for our arguments, and we delete them.

**Case d.2.1** $x^{i_0}(b_1b_2) = 0$. Consider the links incident with $a$. If none of these links have an end-node in $\{b_1, b_2\}$, then all of these links have their other end-nodes in $\mathcal{R}(T_v)$, hence, $ticket(T_v)$ w.r.t. $x^{i_0}$ is $\geq \frac{1}{2}$, and this is a contradiction.

Suppose that $x^{i_0}(ab_2) = 1$. Recall that both $a$ and $b_2$ are "leaf descendants" of the node $q$, which is a descendant of $v$. Note that no other link has an end-node at $b_2$, since $x^{i_0}(\delta_E(b_2)) = 1$ by Corollary 4.5. Consider the links covering the tree-edge between $v$ and its parent. The sum of the $x^{i_0}$ values of these links is $\geq 1$. If all of these links have an end-node in $\mathcal{R}(T_v)$, then $ticket(T_v)$ w.r.t. $x^{i_0}$ is $\geq \frac{1}{2}$, and we have a contradiction. Thus, one of these links has an end-node at $b_1$, and by integrality of $x^{i_0}$ on $J$, the $x^{i_0}$ value of this link is 1. This implies that no other link has an end-node at $b_1$. Now, consider the links covering the tree-edge between $q$ and its parent. The sum of the $x^{i_0}$ values of these links is $\geq 1$. It can be seen that each of these links has an end-node in $\mathcal{R}(T_v)$, because none has an end-node in $\{b_1, b_2\}$ (even if one of these links has an end-node at $a$, the other end-node must be in $\mathcal{R}(T_v)$). Then $ticket(T_v)$ w.r.t. $x^{i_0}$ is $\geq \frac{1}{2}$, and we have a contradiction.

Thus, we must have $x^{i_0}(ab_1) = 1$. Consider the links covering the tree-edge between $v$ and its parent. The sum of the $x^{i_0}$ values of these links is $\geq 1$. If all of these links have an end-node in $\mathcal{R}(T_v)$, then $ticket(T_v)$ w.r.t. $x^{i_0}$ is $\geq \frac{1}{2}$, and we have a contradiction. Thus, one of these links has an end-node at $b_2$. Then it can be seen that $T_v$ satisfies the conditions for a deficient 3-leaf configuration (Figure 1(a) of [6]).

**Case d.2.2** $x^{i_0}(b_1b_2) = 1$. Consider the links incident with $a$. None of these links have an end-node in $\{b_1, b_2\}$ because $x^{i_0}(\delta_E(b_1)) = 1 = x^{i_0}(\delta_E(b_2))$ by Corollary 4.5. Thus, all of these links have their other end-nodes in $\mathcal{R}(T_v)$, hence, $ticket(T_v)$ w.r.t. $x^{i_0}$ is $\geq \frac{1}{2}$; this is a contradiction. □

Dept. of Comb. & Opt., University of Waterloo, Waterloo, Ontario N2L3G1, Canada.
*E-mail address*: jcheriyan@uwaterloo.ca

*E-mail address*: z9gao@uwaterloo.ca